**PROCEEDINGS A**

royalsocietypublishing.org/journal/rspa

## Research

Article submitted to journal

**Subject Areas:**
quantum physics, quantum engineering, computer modelling and simulation

**Keywords:**
Optical atomic clock, single ion trap, space clock, laser cooling, strontium ion, finite element analysis
# An ion trap design for a space-deployable strontium-ion optical clock

A. Spampinato, J. Stacey, S. Mulholland, B. I. Robertson, H. A. Klein, G. Huang, G. P. Barwood, P. Gill

National Physical Laboratory, Hampton Road, Teddington, Middlesex, TW11 0LW, United Kingdom**Author for correspondence:**
Alessio Spampinato
e-mail: alessio.spampinato@npl.co.ukOptical atomic clocks demonstrate a better stability and lower systematic uncertainty than the highest performance microwave atomic clocks. However, the best performing optical clocks have a large footprint in a laboratory environment and require specialist skills to maintain continuous operation. Growing and evolving needs across several sectors are increasing the demand for compact robust and portable devices at this capability level. In this paper we discuss the design of a physics package for a compact laser-cooled $^{88}$Sr$^+$ optical clock that would, with further development, be suitable for space deployment. We review the design parameters to target a relative frequency uncertainty at the low parts in $10^{18}$ with this system. We then explain the results of finite element modelling to simulate the response of the ion trap and vacuum chamber to vibration, shock and thermal conditions expected during launch and space deployment. Additionally, an electrostatic model has been developed to investigate the relationship between the ion trap geometrical tolerances and the trapping efficiency. We present the results from these analyses that have led to the design of a more robust prototype ready for experimental testing.**THE ROYAL SOCIETY PUBLISHING**

©2023 The Author(s) Published by the Royal Society.

# 1. Introduction

There is considerable international activity in the development of optical clocks and their intercomparison with the drive towards a future redefinition of the second [1]. The roadmap towards a redefinition [2] includes targets based on the recommendations in Reference [3]. These include comparisons between optical clocks with an uncertainty of better than 5 parts in $10^{18}$ and target uncertainties for individual devices of 2 parts in $10^{18}$. Optical clocks based on neutral atoms confined in an optical lattice trap allow intercomparison at this level with frequency instabilities at the $10^{-18}$ level averaged over an hour [4, 5]. However, these are complex systems that comprise a significantly higher size, weight and power (SWaP) profile. Whilst single ion trap optical clocks have a greater instability of typically $10^{-15}/\sqrt{\tau}$ over an averaging time $\tau$, they can more easily be developed as compact and robust portable devices [6]. In this paper we present our design for a robust $^{88}Sr^+$ trap package targeting a $5 \times 10^{-18}$ noise floor and $3 \times 10^{-15}/\sqrt{\tau}$ instability that can target applications both in space and on ground.

In addition to applications in ground-based optical frequency metrology, a compact robust optical clock in space opens new opportunities and increased precision across a range of physics and technology applications, including space-based detection of dark matter [7, 8], relativistic geodesy [9, 10], and enhanced satellite navigation accuracy [11]. As discussed in [12], space-borne optical atomic clocks offer significant opportunities to explore different topics in fundamental physics. They could be used to perform measurements of the gravitational redshift with unprecedented precision, searching for violations of the Einstein Equivalence Principle and detection of gravitational waves from a variety of astronomical entities. Optical atomic clocks can find application in future global satellite navigation and timing systems, with optical clocks in both ground and space segments offering both resilience and enhanced performance with respect to current systems. As discussed in [13], when considering the advances in clock accuracies and stabilities, Earth gravity fluctuations (e.g. tides and seismic noise) introduce instabilities that can limit the measurement of time on the surface of Earth. By locating clocks in medium-Earth (MEO) and high Earth (HEO) orbits, it is possible to reduce sensitivity to Earth tidal motion/gravitational noise as well as the gravitational red shift, opening up the idea of future space-based master reference clocks.

The single trapped $^{88}Sr^+$ ion system is an optical secondary representation of the second [14], and also a good candidate for a robust, compact, and portable optical clock. The choice of ion is determined partly by the availability of, for example, distributed feedback (DFB) lasers at specific wavelengths, which for $Sr^+$ are needed for ionisation, laser cooling, clock state preparation of the ion and probing the 674 nm optical clock transition in the ion. This quadrupole clock transition has a frequency of ~ 445 THz, with a natural linewidth of 0.4 Hz. DFB lasers are ideal for ion cooling, interrogation and photoionisation, as they have a large mode-hop free tuning range and can be packaged in a compact hermetically sealed casing with an integral thermoelectric cooler. $^{88}Sr^+$ is under study by different groups, including the UK National Physical Laboratory (NPL) where we have already demonstrated frequency agreement between two single $Sr^+$ clock systems at the low parts in $10^{17}$ level [15]. For $^{88}Sr^+$, light for photoionisation [16] at 461 nm can be produced by frequency doubling of a distributed feedback (DFB) laser at 922 nm [17]. A second laser is also needed at 405 nm, which can be multimode, as sold for the blue disc market. Of the NPL lasers for cooling and clear-out (Figure 1), those at 1092 nm and 1033 nm are custom-made DFB lasers where the frequency drift is corrected using a wavemeter [18]. Therefore, at NPL, only the cooling and clock lasers remain as extended cavity lasers. With further technological development, we anticipate that DFB lasers could also become available at 674 nm and at either 844 nm (doubled to 422 nm) or directly from a GaN laser. Together with an upgraded NPL trap based on [19], the single ion system could ultimately be developed as a robust and continuously operating optical clock with an uncertainty in the low parts in $10^{18}$ region, targeting both future space deployment and ground segment timing applications as well as optical frequency metrology support for a future redefinition of the second.

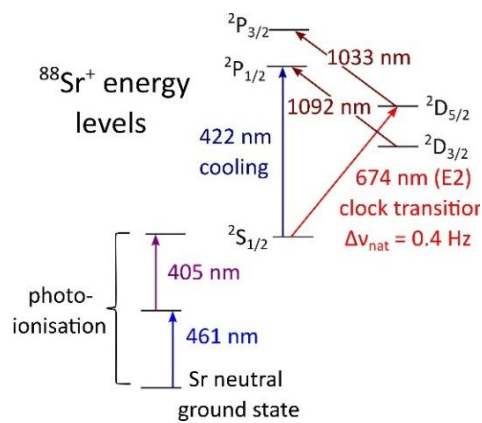

**Figure 1: Partial term scheme for $^{88}Sr^+$ showing the cooling, clear-out and clock transitions. Also shown are the neutral strontium transitions involved in photoionisation**

The $^{88}$Sr$^+$ clock has an estimated quantum projection noise (QPN) instability limit of $2.3 \times 10^{-15}/\sqrt{\tau}$ with $3 \times 10^{-15}/\sqrt{\tau}$ already demonstrated at the National Research Council in Canada [20]. Work on $^{88}$Sr$^+$ is also underway in Finland, using a trap based on the NPL design [21]. We note that an instability of $3 \times 10^{-15}/\sqrt{\tau}$ will reach a noise floor of $5 \times 10^{-18}$ in less than 4 days.

Our design aims to improve functionality and manufacturability. We plan to evaluate the design by constructing a laboratory demonstrator. In validating the physics package for space, the design was subjected to virtual environmental loads including shock and vibration, simulating the response to the mechanical loads associated with launch vehicles and orbit deployment. Electrostatic models were developed to investigate the relationship between the ion trap geometry and the trapping potential. Investigating the effect of misalignments on the trapping potential allowed us to validate the assembly tolerances. Additionally, thermal models were used to characterise the thermal response of the system. The latter are in support of planned thermal vacuum tests, representing a first step towards the development of an engineering model (i.e., a design to be used in a qualification test programme). The thermal models were also used to improve the design of the atomic mini oven by reducing power dissipation, thus allowing shorter turn-on and turn-off times thereby reducing UHV pressure rises during the single ion loading period at start up. In the following sections, we present our latest trap design, followed by a discussion of the structural, electrostatic, and thermal models.

## 2. Physics package design

In this section we review the overall trap design, including the parameters required to meet our target frequency stability and reproducibility. An endcap electrode geometry [22] is most commonly used for single ion optical clocks as it provides a compromise of a largely quadrupolar trapping potential and also an open electrode structure for efficient fluorescence detection. The trapping potential is most straightforwardly expressed in spherical polar coordinates (with the origin at the ion location) and the electrode symmetry means that the potential should be independent of the azimuthal angle $\varphi$. The trap potential $\Phi$ satisfies $\nabla^2 \Phi = 0$ and, for solutions finite at the origin (Figure 2), can be written as a series expansion:

$$\Phi(r,\theta) = \sum_{n=1}^{\infty} a_n P_{2n}(\cos\theta) r^{2n}. \qquad (2.1)$$

Here, $P_n$ are Legendre polynomials of order $n$, $\{a_n\}$ are constants determined by the boundary conditions, $r$ the radial distance and $\theta$ the polar angle. Only terms involving $2n$ are included because the trap electrode symmetry (as shown in Figure 2) yields $\Phi(r,\theta) \equiv \Phi(-r,\theta)$. As explained in [22], hyperbolic electrodes give a solution that only involves $n = 1$, but the requirement of a central electrode ring in this case obstructs much of the fluorescence field of view and decreases optical access. Variants of the open end-cap trap design [22] as implemented, for example, in our previous trap [19] have a smaller quadrupolar field component (i.e., the first term of the above series expansion) and higher order terms are needed when calculating the potential away from the origin.

All aspects of the trap design are considered to reduce/eliminate the causes of potential frequency shifts/systematic errors in the clock operation. To determine the evolution of the ion's position, $r$, we write the ion's equation of motion as $m\ddot{r} = eE$, with $m$ the mass of the ion and $e$ is the elementary charge. The electric field is $E$ found by the gradient of the electric potential $E = -\nabla\Phi$. Static electric fields cannot confine ions in three dimensions and can only produce saddle points, trapping in two directions but with the ion repelled along the third axis. A radio frequency (RF) ion trap overcomes this limitation by using oscillating fields. The resulting time-averaged pseudo-potential can stably trap ions in all three dimensions. As shown in Figure 2, the trap consists in a pair of central electrodes (pin-shaped) surrounded by a pair of hollow electrodes. A gap is present between the electrodes and the electrical separation between these two sets is achieved using ceramic spacers that mechanically connects them to the rest of the trap. An RF voltage is applied to the inner electrodes and a DC voltage applied to the outer electrode. In the equation above, this means that the set of constants $\{a_n\}$ is a sum of both DC and RF components. The ion motion is then described by a Mathieu equation (see, for example [23]).

With reference to the trap schematic in Figure 2, the DC and RF electrodes are installed on a C-shaped component, which is mounted on a support structure (RF feedthrough) exploiting tight tolerances and alignment features. The feedthrough is connected to titanium flange through an alumina disc that provides electrical insulation. As in the baseline design [19], the RF electrode tips have a diameter of 750 μm and a separation of 1 mm. The symmetric "C" component means that the RF drive to the two end caps does not have any significant differential phase shift, so that micromotion [24] can be reduced to sufficiently low levels for an ion at the RF trap centre. Both the RF feedthrough, the electrodes and the "C" component will be made of molybdenum, due to a high electrical and thermal conductivity, together with a low outgassing rate that will reduce ion loss via chemical reactions of the ion with reactive gasses such as oxygen. Molybdenum has also a low coefficient of thermal expansion, is non-magnetic and can be polished to a high surface quality. The DC electric field in the trapping region will be controlled by applying static voltages to the DC electrodes and to four molybdenum compensation electrodes.

Regarding the size of the physics package (excluding the electronics, which are not part of this study), this is dictated by the custom titanium vacuum chamber containing the ion trap assembly. Its approximate dimensions are (L) 130 mm x (W) 45 mm x (H) 150 mm, accounting for the ConFlat® (CF) flanges used for the installation of viewports and feedthrough components. It is worth highlighting that the size of physics package will be reduced during further development activities leading to an engineering qualification model. Parallel experience with other projects at NPL may also allow implementing ion trap physics package design solutions leading to a lower SWaP profile.

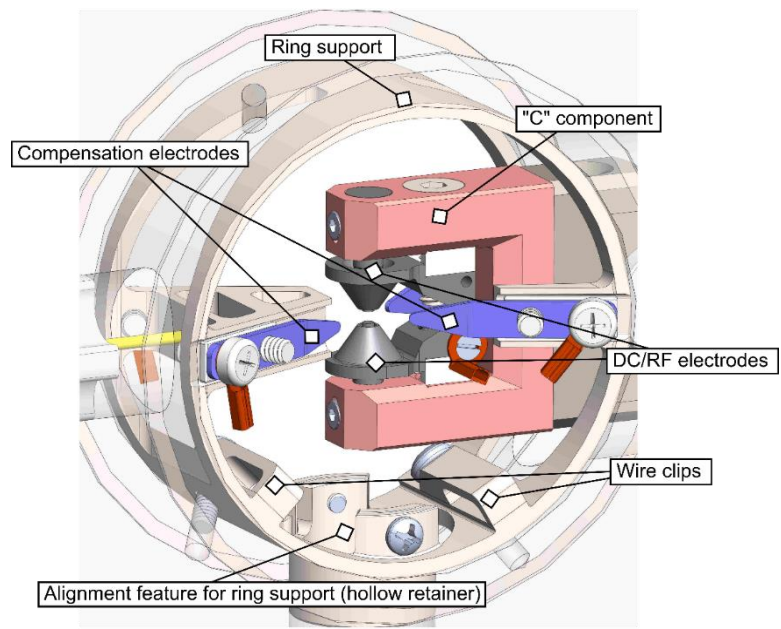

**Figure 2.** Ion trap assembly with key features. The figure shows both structural (ring support, "C" component) and functional (electrodes, wire clips and hollow retainer) elements included in the final design iteration

In our trap, the choice of RF drive frequency is set to exploit the property of some ion clocks where the Doppler and Stark shifts resulting from ion RF motional (Doppler) and electric field (Stark) effects respectively are of opposite sign. For a trap operating at a specific "magic" RF drive frequency, these shifts will therefore cancel. For example, this drive frequency is 25.54 MHz in $^{40}$Ca$^+$ [25] and 14.4 MHz in $^{88}$Sr$^+$ [26]. Ion micromotion [24] at the drive frequency will be minimised by positioning the ion at the RF trap centre via adjustable DC electrode voltages. Finite Element Analysis (FEA) modelling (section 3) is used to calculate the temperature increase in the trap structure resulting from dielectric loss in the insulators which isolate the RF and ground sections [27]. It is important to select materials with a low loss tangent and relative permittivity to minimise thermal heating [19]. In the assessment of $^{88}$Sr$^+$ systematics in Reference [26], the blackbody shift at $T$ = 300 K is determined as 0.24799(37) Hz varying approximately as $T^4$. To keep the blackbody relative frequency shift to 10$^{-18}$, we need to reduce variations in the ion thermal environment, averaged over $4\pi$ steradians, to below 0.14 K. It is anticipated that the spacecraft platform thermal control together with active and passive ion trap package thermal control will enable operational temperatures experienced by the ion within the vacuum chamber to be close to 300 K.

Finally, we note other parameters that will impact our $3 \times 10^{-15}/\sqrt{\tau}$ instability and $5 \times 10^{-18}$ uncertainty targets. These include the allowable phonon heating rates [28, 29, 30] which can be compared with those measured in [19]. Magnetic field noise [31] above a certain level will increase the observed frequency instability and cause variations in the quadrupole shift [32]. A similar fluorescence rate collection efficiency to that previously used [19] will be employed to allow a demonstration of frequency instability close to the quantum projection noise limit. Our trap base pressure $(< 10^{-10}$ mbar) also needs to be sufficiently low to minimise pressure-related relative frequency shifts to the 10$^{-18}$ level [33].

## 3. Structural analyses for space deployment

The launch vehicle and space deployment conditions encountered in European Space Agency missions lead to constraints on the maximum loads (both static and dynamic) that will be experienced by the payload. For this study, a potential mission scenario was used to define the mechanical load environment acting on the physics package during the launch phase. This mission would aim at studying fundamental physics concepts such as gravitational waves, precision tests of general relativity and search of dark matter. The equipment would either be launched as part of a large platform (e.g. mass ~2000 kg) or as a smaller satellite as a secondary payload. This would be positioned on a heliocentric orbit at the Lagrange point L1. Nevertheless, other orbits of scientific interest could be targeted. Such mission concept allowed a down selection of the potential launch vehicles, eventually leading to the Falcon 9 launcher from SpaceX. The expected nominal mechanical loads associated with this launch vehicle [34]) were used for the FEA conducted on the ion trap design, in terms of quasi-static acceleration, sinusoidal vibration, random vibration and mechanical shock. In fact, specific mechanical loads can be associated with the different stages of a launch into space: quasi-static acceleration can be directly linked to the acceleration imposed by the launcher during ascent to orbit; sinusoidal vibration loads are caused by phenomena like unstable combustion, and imbalances in rotating equipment (e.g. turbopumps); random vibrations are linked to acoustic fields generated during lift-off and aerodynamic excitations; mechanical shock loads are due to events like fairing separation, booster burn out, stage separation, solar panel deployment and payload release. The load analyses were conducted using COMSOL Multiphysics on a computer aided design (CAD) model representing the trapped ion space atomic clock assembly. The structural analyses simulated part of the qualification process for space deployment (Figure 3), aiming at providing evidence that the ion trap will withstand a space launch with qualification factors (KQ) applied to the nominal load values. The load qualification values used in the analyses (results in sections (a) to (e)) were obtained in accordance with ECSS-E-ST-10-03C [35], with the qualification factors defined in ECSS-E-ST-32-10C [36].

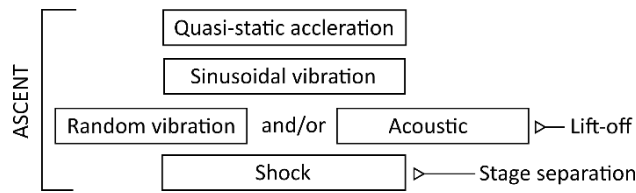

**Figure 3.** Sub-set of mechanical loads for qualification considered for this study (extract of ECSS-ST-10-32C [36]), linked to a corresponding phase/event during launch

For all the developed FEA models, pre-processing operations on the geometry, including defeaturing and meshing, (Figure 4) were required to minimise computing time. In the following sections, the assembly's reference system (Figure 4a) will be used to indicate the different directions of the applied loads; this is a Cartesian reference system, with directions X, Y and Z.

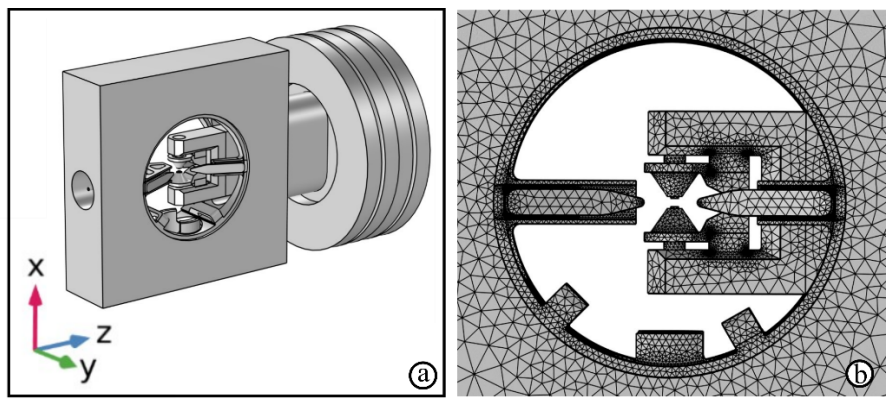

**Figure 4.** Defeaturing (a) and meshing operation (b). The Cartesian reference system associated with the assembly is shown in (a)

## (a) Static analysis

An analysis of the stress/deflection caused by quasi-static acceleration on the ion trap assembly was carried out using the design load factors corresponding to the launch vehicle. These are provided in the table below in terms of axial and lateral acceleration (with respect to the launcher longitudinal axis and in accordance with the user manual [34]).

**Table 1.** Flight limit loads associated with a Falcon 9 launch vehicle (nominal expected loads and qualification values)

| | Axial Acceleration [g] | | Lateral Acceleration [g] | |
|---|---|---|---|---|
| Point in the static load envelope, Falcon User's guide [34] | Nominal | Qualification | Nominal | Qualification |
| A | 6.0 | 7.5 | 0.5 | 0.6 |
| B | 4.0 | 5.0 | 0.5 | 0.6 |
| C | 3.5 | 4.5 | 2.0 | 2.5 |
| D | -1.5 | -1.9 | 2.0 | 2.5 |
| E | -1.5 | -1.9 | 0.5 | 0.6 |
| F | -2.0 | -2.5 | 0.5 | 0.6 |

To simulate the largest expected accelerations, the maximum values for the axial acceleration (7.5 g – Point A in Table 1) was combined with the maximum lateral acceleration (2.5 g – Point C or D in Table 1). Preliminary analysis of the quasi-static acceleration carried out during several iterations leading to the final design shown that the most critical condition for the ion trap components (in terms of stress and deflection) corresponds to the case of an axial load applied in the Y direction, and a lateral load applied in the X direction.

The maximum equivalent stress (von Mises, $S_{max}$; see Figure 5a) was found on the support ring and does not exceed 1.07 MPa. Considering a yield design factor of safety (FOSY) of 1.25 for metallic parts ( [36], verification by analysis only) and as the material applied to this component was titanium with an assumed yield strength ($S_{yield}$) of 880 MPa, the analysis demonstrated through a large margin of safety (MoS = ($S_{yield}$ / FOSY · $S_{max}$) - 1 = 657) that the applied load does not lead to plastic deformation or fracture of this component.

Additionally, the analysis of the predicted displacement (Figure 5b) showed that the maximum deflection for the ion trap components corresponds to the ring support (0.24 µm). The simulated deflection values are much smaller than the clearance between components, suggesting that there is no risk of these clashing due to quasi-static acceleration.

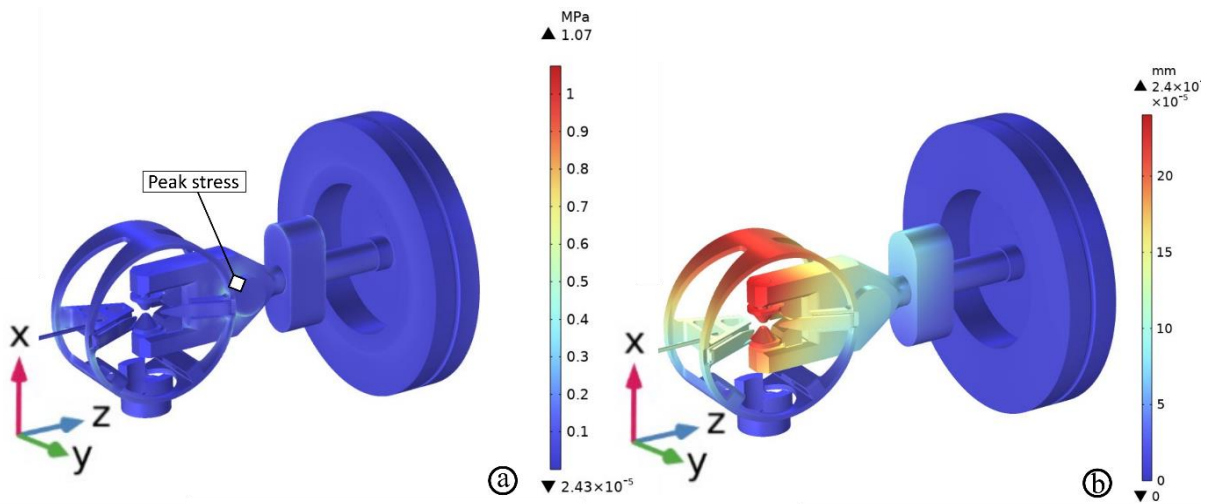

**Figure 5.** Maximum equivalent stress (a) and maximum deflection (b) for the ion trap key components

### (b) Modal analysis

A modal analysis was conducted on the ion trap assembly, obtaining the system's eigenfrequencies and the related eigenmodes, forming the bases for a set of dynamic analyses (sinusoidal acceleration, random vibration and shock). The relative modal mass contribution in the three orthogonal directions of the 20 lowest frequency modes is shown in Table 2 as the percentage of the ion trap sub-assembly mass that participates in a particular mode. In Table 2, the first 6 modes with a value of the relative modal mass contribution in either of the three directions greater than 5% are highlighted, as these modes represent the most significant contributors to the dynamic response of a system. The values shown in the table represent a subset of the 120 modes found with this analysis, necessary to include the largest possible fraction of the system's total mass for the correct execution of the dynamic analyses.

Table 2. Ion trap assembly eigenfrequency and mass participation factors

| Mode | Frequency [Hz] | Relative modal mass contribution, X-direction | Relative modal mass contribution, Y-direction | Relative modal mass contribution, Z-direction |
|---|---|---|---|---|
| 1 | 3006.8 | 0.00% | 27.08% | 0.00% |
| 2 | 5610.8 | 0.00% | 1.67% | 0.00% |
| 3 | 6608.9 | 4.50% | 0.00% | 11.73% |
| 4 | 7306.1 | 1.11% | 0.00% | 22.12% |
| 5 | 7313.5 | 0.00% | 0.58% | 0.07% |
| 6 | 7738.4 | 0.00% | 0.00% | 0.03% |
| 7 | 7794.8 | 0.00% | 0.00% | 0.02% |
| 8 | 10890 | 0.00% | 0.16% | 0.00% |
| 9 | 11094 | 20.47% | 0.00% | 2.78% |
| 10 | 11634 | 0.81% | 0.00% | 0.00% |
| 11 | 11802 | 0.67% | 0.00% | 0.80% |
| 12 | 13092 | 0.00% | 1.19% | 0.00% |
| 13 | 13685 | 0.02% | 0.00% | 0.09% |
| 14 | 14091 | 0.00% | 10.57% | 0.00% |
| 15 | 14948 | 0.03% | 0.00% | 0.00% |
| 16 | 16140 | 7.69% | 0.00% | 4.31% |
| 17 | 16666 | 0.00% | 0.03% | 0.00% |
| 18 | 16707 | 2.91% | 0.00% | 0.85% |
| 19 | 17307 | 0.00% | 1.78% | 0.00% |
| 20 | 17359 | 0.27% | 0.00% | 0.12% |

Figure 6 shows the mode shapes (deflection patterns associated with a particular modal frequency) for the ion trap sub-assembly corresponding to the 6 lowest eigenfrequencies associated with a relative modal mass contribution greater than 5% (from Table 2). This figure highlights the components with the larger contribution to the mode, which is related to the resultant deflection linked to a potential resonance. Each of the modes displayed in Figure 6 involves one or more components in the sub-assembly, with different characteristic responses of the system in case one of those frequencies is excited by an external load. It is worth highlighting that, as the lowest eigenfrequency is ~3000 Hz, only mechanical shock loads could cause a resonance of this sub-system, due to these being the only loads acting above 2000 Hz (in the range 100-10000 Hz – as discussed in section (e)). The modes shown in Figure 6 are mainly associated with the DC/RF electrodes, the support ring and the compensation electrodes, corresponding to translations in the XZ and YZ planes.

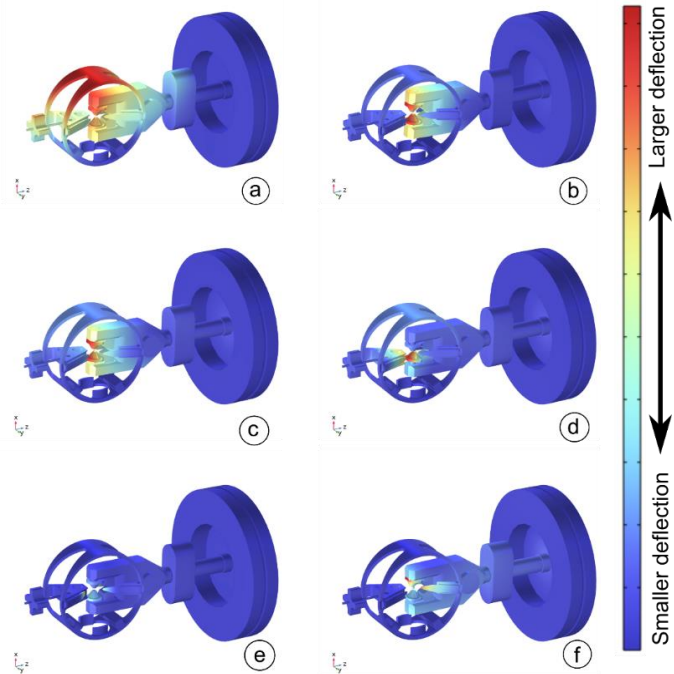

**Figure 6.** Modes of the ion trap sub-assembly for the first 6 modes with a relative modal mass contribution >5%: (a) 3006.8 Hz, (b) 6608.9 Hz, (c) 7306.1 Hz, (d) 11094 Hz, (e) 14091 Hz, (f) 16140 Hz.

## (c) Harmonic response analysis

An analysis of the stress/deflection linked to the sinusoidal vibration environment was carried out using the reported levels at the top of the payload attach fitting. These are presented in Table 3 and Table 4 both for the axial and the lateral acceleration (as before, with respect to the launcher longitudinal axis).

**Table 3.** Axial acceleration values for the limit sinusoidal vibration (nominal expected loads and qualification values)

| Frequency [Hz] | Axial Acceleration [g] - nominal | Axial Acceleration [g] – qualification |
| --- | --- | --- |
| 5 | 0.5 | 0.7 |
| 20 | 0.8 | 1.0 |
| 35 | 0.8 | 1.0 |
| 35 | 0.6 | 0.8 |
| 75 | 0.6 | 0.8 |
| 85 | 0.9 | 1.1 |
| 100 | 0.9 | 1.1 |

**Table 4.** Lateral acceleration values for the limit sinusoidal vibration (nominal expected loads and qualification values)

| Frequency [Hz] | Lateral Acceleration [g] - nominal | Lateral Acceleration [g] - qualification |
| --- | --- | --- |
| 5 | 0.5 | 0.6 |
| 85 | 0.5 | 0.6 |
| 100 | 0.6 | 0.8 |

Preliminary analysis of the quasi-static acceleration carried out during design iterations shown that the most critical condition corresponds to the case of an axial load applied in the Y direction, combined with a lateral load applied in the X-direction.

The maximum equivalent stress corresponds to 0.7 MPa at the highest frequency value for the sinusoidal load, discovered on the molybdenum RF feedthrough rod. As this stress value is lower than what was discovered with the static analysis in section (a), it is concluded that there is sufficient MoS and this load is not likely to lead to a failure (or plastic deformation) scenario. The predicted peak maximum deflection was found on the upper part of the vacuum chamber (0.05 µm), while the ion trap components are subjected to lower deflection values (0.04 µm). As these values are significantly smaller than the dimensions characterising the sub-assembly, the risk of components clashing during a sinusoidal load was deemed to be extremely low.

### (d) Random vibration analysis

A study on the response of the ion trap sub-assembly to random vibration was performed, calculating the resulting Root Mean Square (RMS) acceleration and the Power Spectral Density (PSD) based on the maximum predicted random vibration environment (i.e. vibration with no periodicity) associated with the launch vehicle. The vibration levels are summarised in the table below in terms of PSD, showing both the nominal and the qualification values.

**Table 5.** Random vibration environment linked to the Falcon 9 launch vehicle (nominal expected loads and qualification values)

| Frequency [Hz] | PSD [$g^2$/Hz] - nominal | PSD [$g^2$/Hz] - qualification |
| --- | --- | --- |
| 20 | 0.0044 | 0.0110 |
| 100 | 0.0044 | 0.0110 |
| 300 | 0.01 | 0.03 |
| 700 | 0.01 | 0.03 |
| 800 | 0.03 | 0.08 |
| 925 | 0.03 | 0.08 |
| 2000 | 0.00644 | 0.01610 |

From several analyses conducted to validate the design, the most critical condition in terms of RMS acceleration and resulting PSD was found to be for the case of an excitation applied in the Y direction. Considering the entire assembly, a maximum RMS acceleration of 32.2 m/$s^2$ (~ 3.28 g) was discovered on the top part of the vacuum chamber (Figure 7a), while the maximum RMS acceleration of the ion trap components (Figure 7b) was found to be 21.8 m/$s^2$ (~ 2.22 g) for the ring support; the "C" and the DC/RF electrodes presented similar values. As the RMS acceleration does not give direct information about its peak value, it is common practice to assume that the average one-sided peak level is about three times the RMS value (3σ value). As such, the peak acceleration for the ion trap assembly was estimated to be ~6.7 g and judged to be low enough (in comparison with the most severe quasi-static acceleration values shown in section 3a – Point A) not to pose a risk for the structural integrity of the assembly.

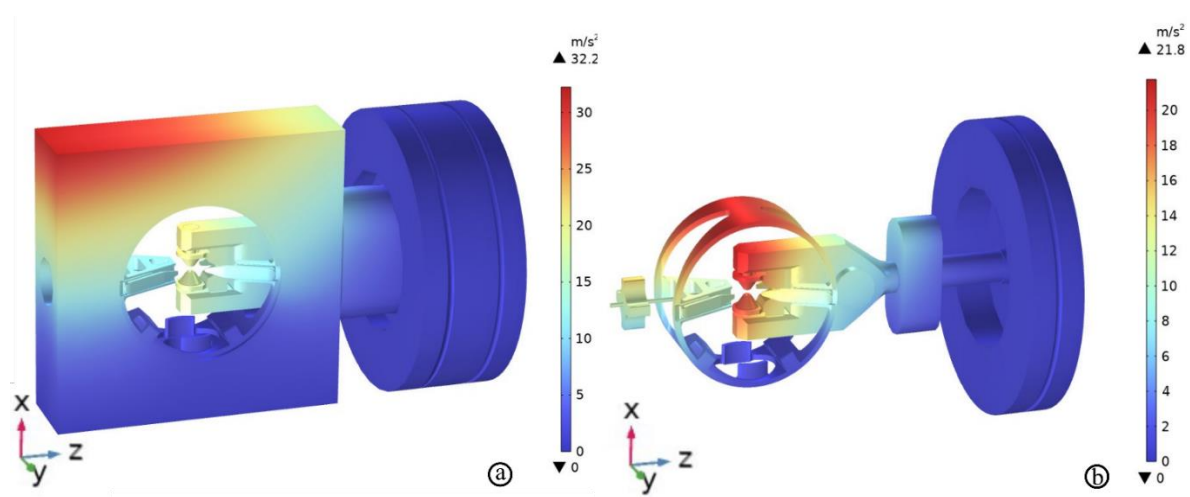

**Figure 7.** RMS acceleration of the ion trap assembly caused by excitation in the Y direction. Peak acceleration on the vacuum chamber (a) and focus on the peak acceleration for the ion trap components (b)

## (e) Shock response spectrum analysis

The analysis of the mechanical shock environment acting on the system during a space launch was based on the maximum expected values presented in the table below, expressed in terms of a Shock Response Spectrum (SRS). As shock tests are performed with no margins to consolidate the shock specification of the space segment equipment [35], the nominal values in Table 6 were used in the shock response analysis without any additional qualification factors.

**Table 6. Shock Response Spectrum** associated to the mechanical shock loads associated to the Falcon 9 launcher

| Frequency [Hz] | SRS [g] |
|---|---|
| 100 | 30 |
| 1000 | 1000 |
| 10000 | 1000 |

The shock response spectrum analysis was performed by applying the shock load along the X direction, as this corresponds to the worst-case scenario discovered during preliminary studies to validate design changes. The maximum equivalent stress related to the ion trap assembly (Figure 8a), was found to be $S_{max}$ = 107 MPa for the molybdenum DC electrodes and "C" component. Considering a yield design factor of safety (FOSY) of 1.25 for metallic parts [36] and assuming a yield strength ($S_{yield}$) of 325 MPa for molybdenum, the analysis demonstrated through sufficient margin of safety (MoS = 1.42) that the applied load does not lead to a plastic deformation or fracture of these components. Additionally, the maximum deflection (norm) for the examined case was found to be ~20 μm (Figure 8b), hence excluding the risk of clashing between components.

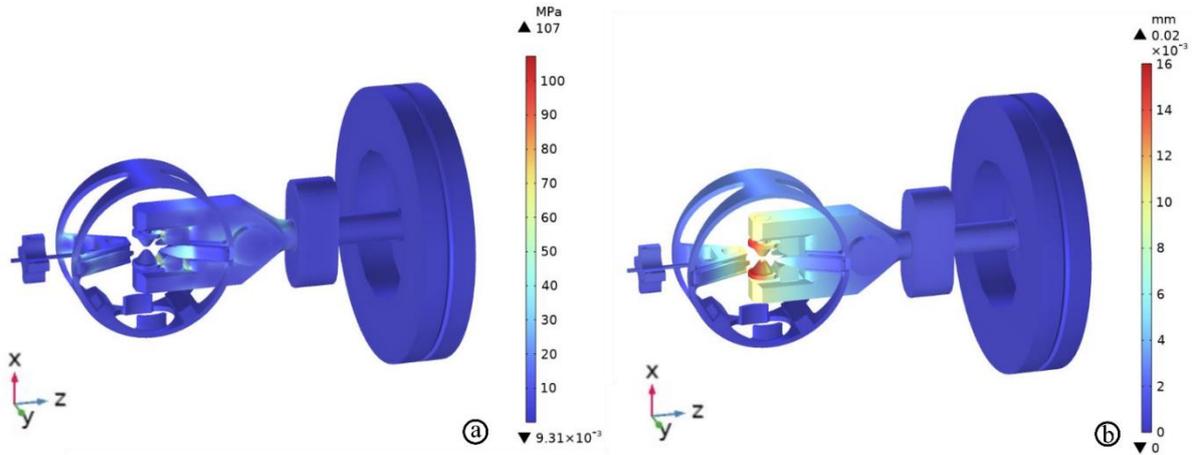

**Figure 8.** Maximum equivalent stress (a) and maximum deflection (b) on the RF feedthrough cantilever caused by a mechanical shock applied in the Y direction

## 4. Electrostatic studies

A series of electrostatic studies on the baseline ion trap design [19] investigated how dimensional variations of the RF and DC electrodes affect the trap functionality in terms of the predicted pseudo-potential $V_{ps}{}^{*}$, calculated through the following equation:

$$V_{ps}{}^{*} = \frac{q^2 \, E^2}{4 m \Omega^2},$$ (4.1)

where $q$ is the ion charge, $E$ is the electric field, $m$ is the ion mass (88 amu) and $\Omega$ is the trap drive angular frequency.

For these studies, an AC amplitude of 300 V was applied to the RF electrodes and the components electrically connected to these (e.g. "C" component, RF feedthrough). An electric potential of 0 V was applied to the DC electrodes and all the electrically grounded elements (e.g. support ring, vacuum chamber).

In addition to the pseudo-potential, the gradient of the electric potential in the region between the trap elements was simulated (section view in Figure 9) and found to match the expected behaviour discussed in [19].

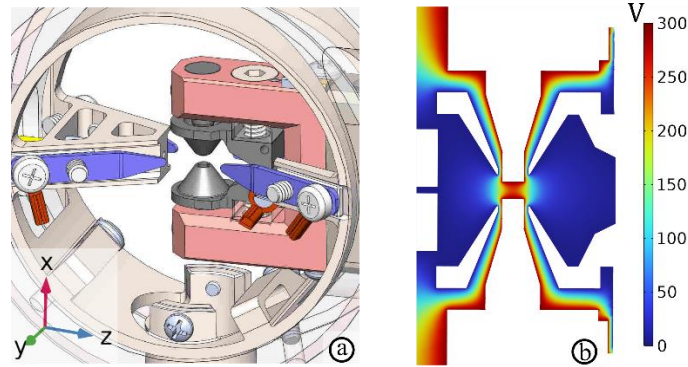

**Figure 9.** Reference system for the ion trap during the electrostatic study (a) and electric potential distribution (b)

With reference to Figure 10, the conducted simulations studied the change in pseudo-potential caused by deviations from the nominal value of the following parameters for the RF electrodes: separation between the electrodes ($d_{RF}$), diameter ($Dia_{RF}$), face angle ($\alpha$) and symmetric misalignment of the upper and lower electrodes ($Mis_{RF}$). Additionally, the effect of changing the separation between the DC electrodes ($d_{DC}$) was studied.

Knowing the effect of these deviations from the baseline design, the electrostatic studies presented in the following sections allowed us to establish the geometrical tolerances to be used in the manufacture of these components. The nominal geometry was found to be optimal (in terms of pseudo-potential fitting characteristics) and based on the outcome of these studies the corresponding manufacturing tolerances were established in the range ± 50 μm (corresponding to the "fine" designation of the DIN ISO 2768-mk standard on manufacturing tolerances).

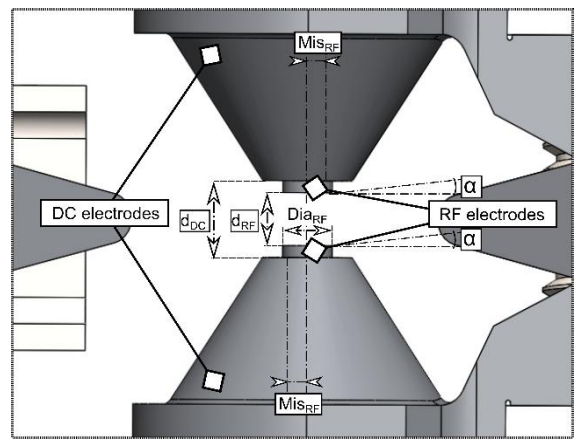

**Figure 10.** Key parameters characterising the ion trap geometry. The parameters in the figure are: separation between the RF electrodes ($d_{RF}$), RF diameter ($Dia_{RF}$), RF face angle ($\alpha$), symmetric misalignment of the upper and lower RF electrodes ($Mis_{RF}$) and separation between the DC electrodes ($d_{DC}$)

The values of the pseudo-potential linked to each of these cases were calculated in the trapping region along the axial (X) and radial (Y, Z) directions with respect to the RF electrodes longitudinal axis. The collected data were then post-processed by fitting a quadratic and extracting the fitting statistics associated to it. Specifically, the Sum of Squares due to Error (SSE) was used as the main indicators of the Goodness-of-Fit, with values closer indicating a better fit. This study provided an indication on how a dimensional change in the electrodes leads to a deviation from the ideal parabolic pseudo-potential for a quadrupole trap (associated with simple harmonic motion).

The following sections discuss the results related to the pseudo-potential calculated in the X direction (axial) of the reference system displayed in Figure 9, as cases representative of separately examined scenarios in the radial directions. The parabolic fitting curves are presented below to demonstrate the change in pseudo-potential with RF electrode separation, which provides a visual indication of the analysis conducted for all the other cases.

### (a) Pseudo-potential as a function of RF electrode separation

For this first scenario, the separation between the RF electrodes was altered, varying this from its initial value (1 mm) in the range [-400 μm (almost touching), +600 μm]. Observing the parabolas in Figure 11, it can be seen that, when compared to the nominal geometry, any variation in the length of the RF electrodes corresponds to a decrease in the pseudo-potential (i.e. less steep parabolas).

Examining the fitting statistics, both the nominal geometry and a change of +200 μm in the electrode separation lead to a good fit of the pseudo-potential parabolic curve. In general, it was noticed that an increased separation of the RF electrodes within the range studied does not lead to a deterioration of the quadratic fitting, while this is not true for a decrease in the separation between these components.

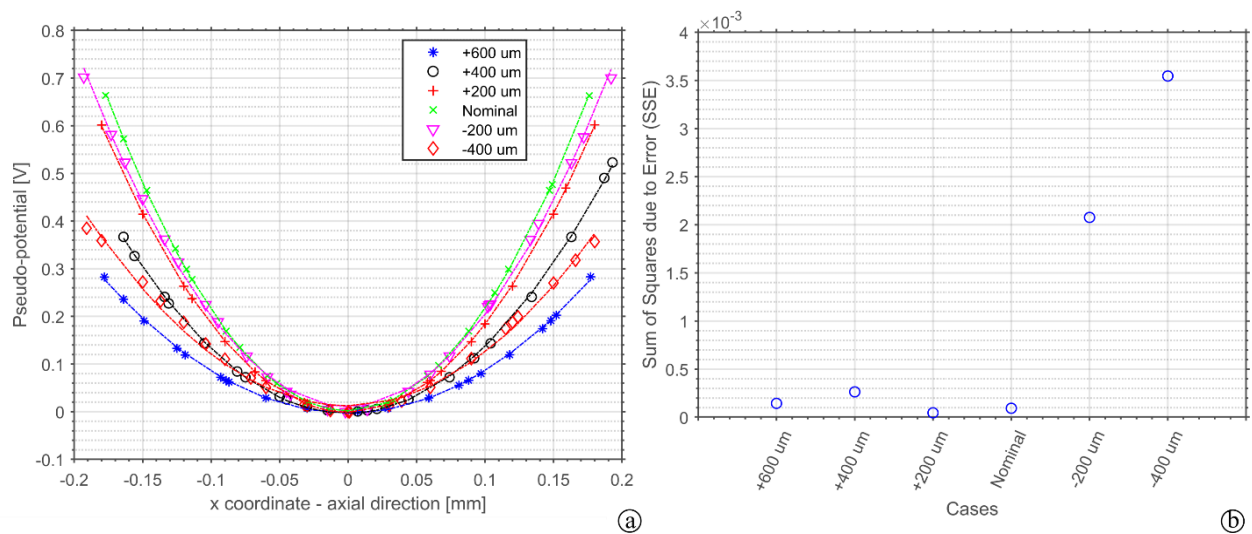

**Figure 11.** Pseudo-potential parabolic fitting curves, (a), and Sum of Squares due to Error, (b), for different RF electrode separations.

### (b) Pseudo-potential as a function of RF electrode diameter

In this second scenario, an electrostatic study was conducted in relation to variations in the RF electrode diameter. The diameters of both electrodes were altered simultaneously, with a deviation in the range [-300 μm, +200 μm] from its nominal value (1 mm). In this instance, an increase in the RF electrodes' diameter corresponds to a decrease in the pseudo-potential (less steep parabolas), while a decrease in the value of this parameter leads to an increase in the pseudo-potential (steeper parabolas). From the fitting statistics, it is seen that both the nominal geometry and a variation of -100 μm in electrode diameter are associated with a good fit of the pseudo-potential parabolic curve. Considering that the difference between the two configurations is minor and considering the achievable manufacturing tolerances (within ±100 μm), this study did not suggest any changes to the nominal trap design.

### (c) Pseudo-potential as a function of RF electrode face angle

To study the effect caused by changes in the angle of the RF electrode face (with respect to a plane parallel to the YZ plane and the central point of the electrode fixed - refer to Figure 9a), this parameter was altered simultaneously for both electrodes in the range [+0.5°, +10°], i.e. they continue to face each other. For this study, large values of this angle (outside any manufacturing tolerances) were purposely considered, to highlight any potential trends in the predicted pseudo-potential. As this operation breaks the rotational symmetry about the X axis, the secular frequencies in the Y and Z directions are no longer equal. As might be expected from symmetry, the nominal geometry scenario corresponds to the best fit of the pseudo-potential parabolic curve. However, we found that the trapping efficiency and harmonicity do not diverge significantly from the nominal case as this angle is varied within the [+0.5°, +10°] range.

### (d) Pseudo-potential as a function of RF electrode misalignment

The effect on the trap's pseudo-potential caused by a misalignment of the RF electrode longitudinal axis (on a plane passing through the nominal axis position and parallel to the XZ plane - refer to Figure 9a) was studied by imposing a change to this parameter to both electrodes simultaneously and in opposite directions. As such, the total misalignment between the electrodes was limited to 300 μm to avoid accounting for scenarios where the RF electrodes overlap with the DC electrodes.

As for the previous face angle change cases, this change breaks the rotational symmetry about the X axis and as a result the secular frequencies in the X and Y directions are no longer equal. A change in this parameter did not lead to any significant changes in the pseudo-potential values with respect to the RF electrode misalignment. We found that the trapping efficiency and harmonicity did not change significantly from the nominal case as this misalignment is increased from 0 to 300 µm which is likely due to variations being small relative to the 750 µm diameter electrode faces.

### (e) Pseudo-potential as a function of DC electrode separation

In the last scenario, the effects on the pseudo-potential caused by a change to the DC electrode separation were examined in the range [-600 µm, +600 µm] from its initial value (1.45 mm). In this case, a decrease in the separation between the DC electrodes leads to an increase in the predicted pseudo-potential and vice versa. Examining the fitting statistics, there was no clear trend in the statistics of the parabolic pseudo-potential fit.

## 5. Thermal analyses

In validating the ion trap design, it was subjected to a series of analyses aiming at studying the thermal behaviour of different parts of the system under specific operational/external conditions. These analyses can be categorised as follows:
- Electro-thermal modelling of the ion trap to simulate the electromagnetic heating of the trap's components linked to the applied RF potential.
- Electro-thermal modelling of the atomic mini oven to study the time-dependent effect of the applied electrical current on the temperature distribution.
- Thermal cycling modelling of the entire system to simulate the conditions encountered during a thermal vacuum test, understanding the system's heating/cooling rates and dwell (time required to ensure that internal parts have achieved thermal equilibrium).

### (a) Electro-thermal modelling of the ion trap

A model was developed to simulate the temperature distribution at steady-state conditions within the ion trap assembly for a specific value of the applied RF electrical potential, due to electromagnetic heating of the trap's components. The model accounts for the electrical properties of the different materials constituting the ion trap.

Boundary and initial conditions were applied to the model to define zero potential surfaces (vacuum chamber and DC electrodes). An RF amplitude of 300 V [19] was applied to the RF electrodes, the "C" component and the feedthrough rod. The initial temperature of the entire assembly was set to the ambient value (20 °C) and a boundary condition was imposed on the vacuum chamber bottom surface to remain at 20 °C. The latter allowed simulating the contact between the vacuum chamber and the breadboard plate, effectively acting as a heat sink. The effect of radiative heat transfer from the ion trap elements towards the surrounding environment was included in the model. Finally, to simulate the effects of applying RF potential to the trap, a frequency of 14 MHz (the expected trap drive frequency, as discussed in [26]) was defined. In this $^{88}$Sr$^+$ trap, we expect this drive frequency at 300 V amplitude to give secular frequencies in the region of 1.5 MHz. With this secular frequency and observed phonon heating rates, we expect to achieve excellent signal contrast with Rabi interrogation of the clock transition. This point is discussed further in [16, 29].

The result of the application of the electric potential are shown in Figure 12a (front view) and Figure 12b (top view) using a sectioning plane passing through the centreline of the RF electrodes and parallel respectively to the XZ and YZ planes. The potential is shown only on the surface resulting from the section operation.

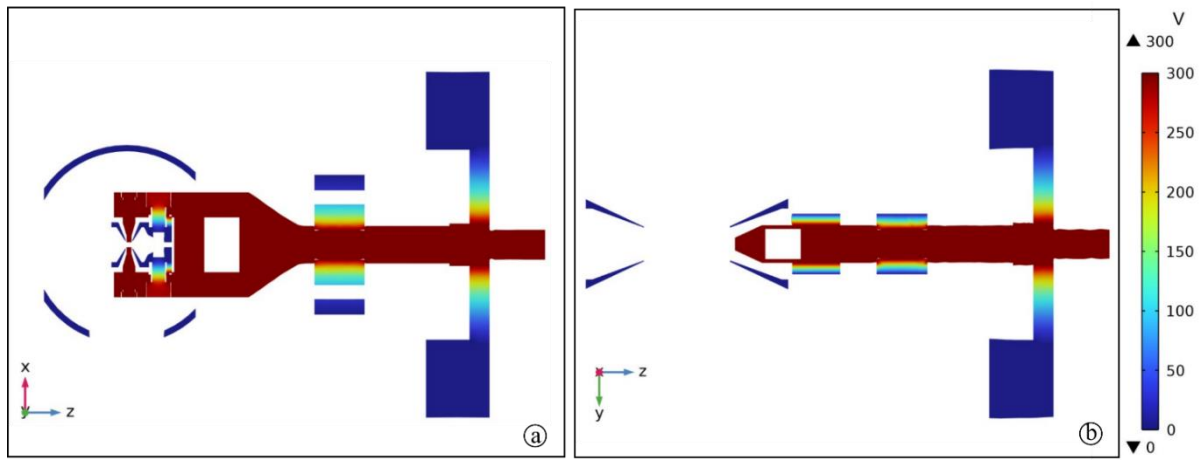

**Figure 12.** Electric potential of the ion trap assembly, section views: front (a) and top (b). The void (white rectangle) in the feedthrough rod is material that has been removed to reduce the mass of the component

A difference of electric potential between the RF electrodes (including "C" and feedthrough rod) and the DC electrodes results in electromagnetic heating of the ion trap components. This effect was simulated in the FEA model and can been seen in Figure 13a (front) and Figure 13b (top). In these figures, section views parallel to the XZ and YZ planes allow for the temperature distribution of the inner features to be observed. A maximum temperature of 20.2 °C is reached on the RF/DC electrodes and supporting "C" components (0.2 °C above ambient). This is in line with the results obtained in previous work focusing on thermal simulation of similar ion trap designs [19].

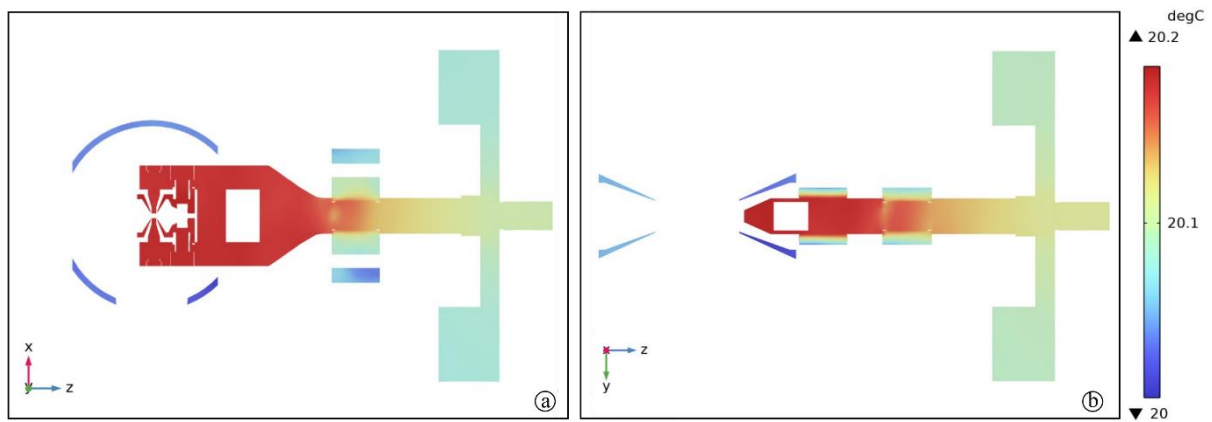

**Figure 13.** Temperature distribution of the ion trap assembly, section views: front (a) and top (b). The void (white rectangle) in the feedthrough rod is material that has been removed to reduce the mass of the component. A thermal boundary condition of 20 °C was imposed on the vacuum chamber (connected to a titanium flange part of the RF feedthrough).

The electro-thermal models of the ion trap verified that the developed ion trap design is expected to heat by 0.2 °C above ambient at its hottest point during operation. This results in the characteristic temperature of the black-body radiation seen by the ion, when averaged over the full $4\pi$ field of view, rising by less than 0.14 °C. The relative frequency shift from this black-body radiation is below $1 \times 10^{-18}$ [26].

## (b) Electro-thermal modelling of the atomic mini oven

An electro-thermal model was created to simulate the temperature distribution in the atomic mini-oven sub-assembly (Figure 14) for a specific value of the electrical current applied to the terminals. This simulation is aimed at understanding the temperature change in different areas of the oven as a function of time, following the application of an electrical current. As such, the time required to reach a specific temperature from an initial ambient value was obtained, together with the rate of temperature increase. Similarly, the time required to return to the ambient temperature and the related decrease rate was predicted. The atomic mini oven is the source of strontium atoms that are then ionized by laser light directed into the ion trap. A tube containing strontium is resistively heated to a temperature where this element begins to emit vapour. As shown in Figure 14 (which also indicates the component materials), the 3D model part of this trap design consists of an oven tube connected to two electrical wires (and related terminals), installed into the support ring within the vacuum chamber and electrically insulated from the rest of the metal chamber components using a ceramic insulator (alumina tube).

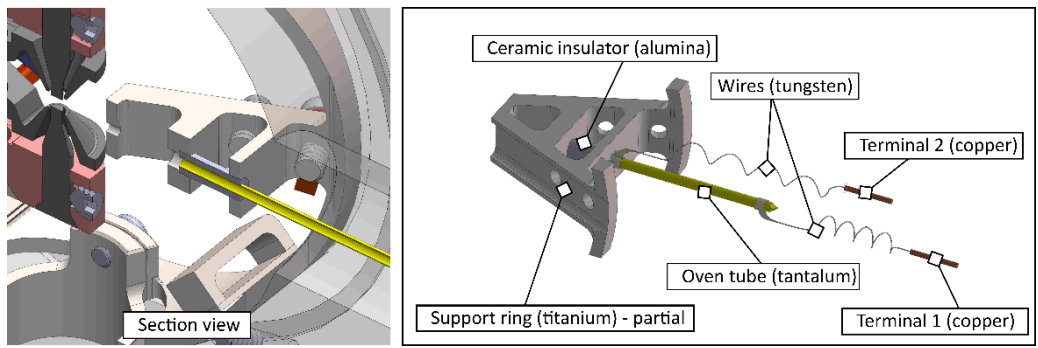

**Figure 14.** Details of the atomic mini oven sub-assembly

Different versions of the oven were modelled, aiming at minimising the input electric current and obtaining a faster heating rate for a target peak oven temperature. The oven models studied here (Figure 15) are described as follows:
i. A version with both electrical wires having a relatively large diameter of 1 mm.
ii. A version with electrical wires having a 0.1 mm diameter. This model was created to understand the effect of a lower wire resistance on the temperature profile within the oven sub-assembly and the required electrical current.
iii. A version with mixed wire diameters (1 mm and 0.1 mm), to verify whether the oven temperature profile and heating rate would benefit from introducing different wire resistance values.

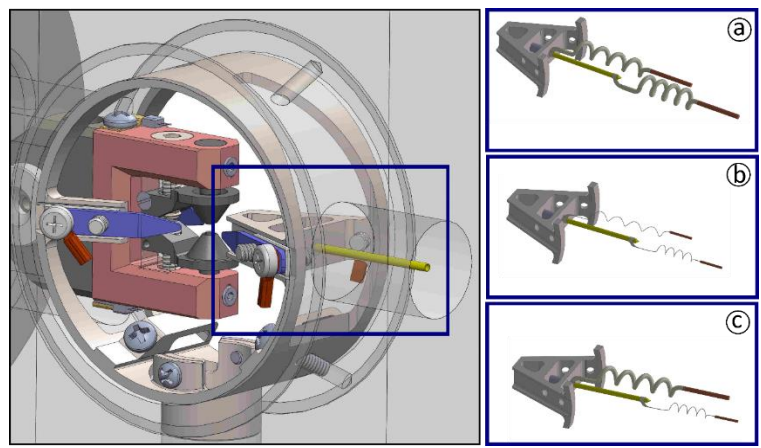

**Figure 15.** Oven configurations modelled for electro-thermal FE analysis: a) thick wires (1 mm in diameter), b) thin wires (0.1 mm in diameter), c) both thick and thin wires

Using a similar approach to the one described in section (a), an electric current was applied to the terminals (Figure 14), simulating the temperature distribution resulting from electromagnetic heating and accounting for the heat dissipation related to the surface-to-ambient radiation. The input current was defined as a function of time, initially set to a constant value and subsequently decreased to zero after having reached steady-state conditions. The electrical current was set to a specific value for each of the model versions in relation to the required peak oven temperature of 300 °C, which gives a high enough vapour pressure of strontium to load the trap [37]. Boundary and initial conditions were applied to the model, defining electrically grounded surfaces (support ring and terminal) and surface ambient temperature (20 °C, for the support ring where in contact with the vacuum chamber and at sectioned features for continuity). Moreover, a contact impedance node was applied, which defined the surface resistance of the of the wire connections to the oven tube (for all versions of the model, this was set to 0.1 Ω). The results from simulating both the temperature distribution and temperature profile in relation to the time steps will be discussed in the following paragraphs, covering three versions of the atomic mini-oven sub-assembly.

### (i) Oven model with 1 mm diameter electrical wires
The first version of the atomic mini oven incorporated coiled electrical wires with a diameter of 1 mm (Figure 16a). In this configuration, a maximum temperature of 300 °C (at steady-state conditions) was reached at the rear end of the oven tube (pinched side, the farthest from the trapping region), using an electric current of 3 A. Both the ring support and the wire connected to the front end of the mini oven showed a temperature in the range 20 °C – 40 °C, while the temperature variation within the oven tube and the wire connected to its rear end was in the range 35 °C – 300 °C. Figure 16b shows the temperature profile as a function of time for a 3 A electric current, for both the rear and front end of the oven tube, together with the average value calculated between with these two areas. The steady-state temperature conditions (300 °C ± 2 °C) for the rear end of the oven tube were reached at around 100 s after applying the current, reaching 90% of the temperature rise after 22 s. Following the drop of current to 0 A, the temperature of the same area of the oven drops exponentially with a $1/e$ decay time of around 80 s.

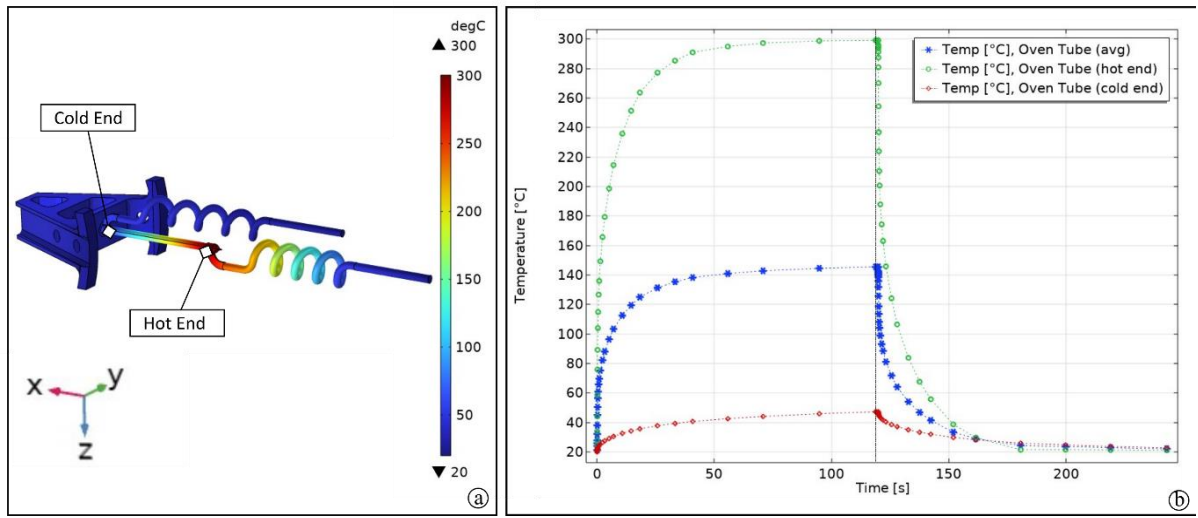

**Figure 16.** 1 mm diameter electrical wires configuration: (a) temperature distribution in the mini oven at the peak heating and (b) temperature profile in relation to the application of electrical current

(ii) Oven model with 0.1 mm diameter electrical wires

The second version of the oven sub-assembly consisted of coiled electrical wires with a diameter of 0.1 mm (Figure 17a). In this configuration, the required target temperature of 300 °C was reached at the rear end of the oven tube applying an electric current of 1.55 A (half the value used in the first case), while a peak temperature value of 888 °C was reached on the wire connected to the rear of the oven. As before, both the ring support and the wire connected to the front end of the mini oven showed a temperature in the range 20 °C – 40 °C, while the temperature variation within the oven tube was in the range 50 °C – 300 °C. Compared with the previous scenario, the steady-state temperature conditions (300 °C ± 2 °C) for the rear end of the oven tube were reached after around 50 s following the application of electric current (Figure 17b). However, following the drop of current to 0 A, a $1/e$ decay time of around 30 s can be observed for the same area of the oven.

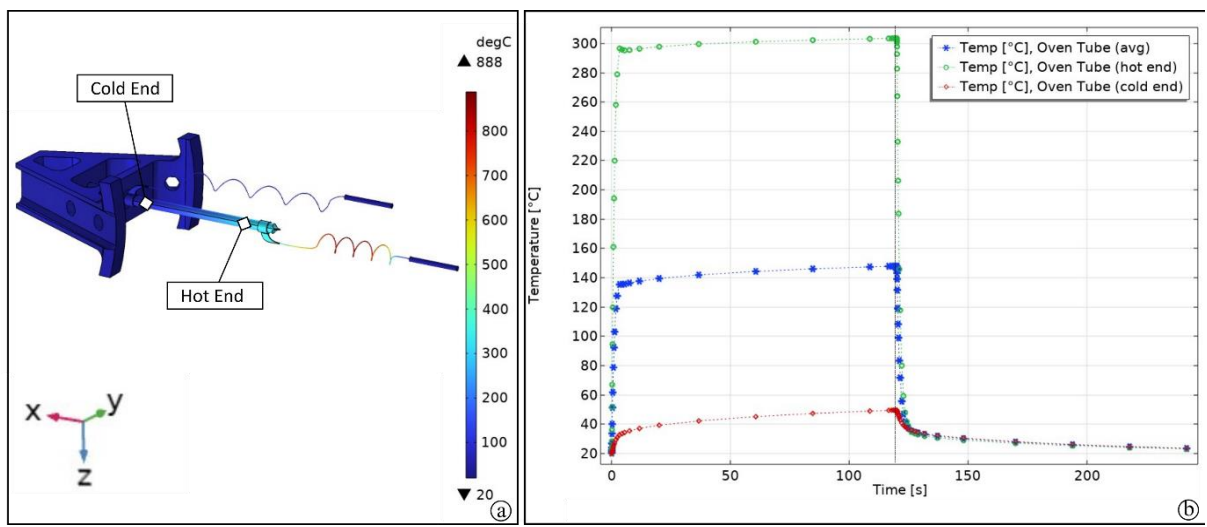

**Figure 17.** 0.1 mm diameter electrical wires configuration: (a) temperature distribution in the mini oven at the peak heating and (b) temperature profile in relation to the application of electrical current

(iii) Oven model with electrical wires of mixed diameters (1 mm and 0.1 mm)

In the third version of the mini oven, coiled electrical wires with of both 1 mm and 0.1 mm diameter. The temperature distribution in the sub-assembly under steady-state conditions is shown in Figure 18a which has similarities to that discussed in the previous scenario, where the required oven target temperature of 300 °C was reached at the rear end of the oven tube applying an electric current of 1.55 A. A peak temperature of 888 °C was reached on the wire connected to the rear of the oven and the same temperature ranges as before were found for the ring support, the electrical wires and the oven tube. The temperature profile in Figure 18b shows the same heat-up and cool-down time rates/times.

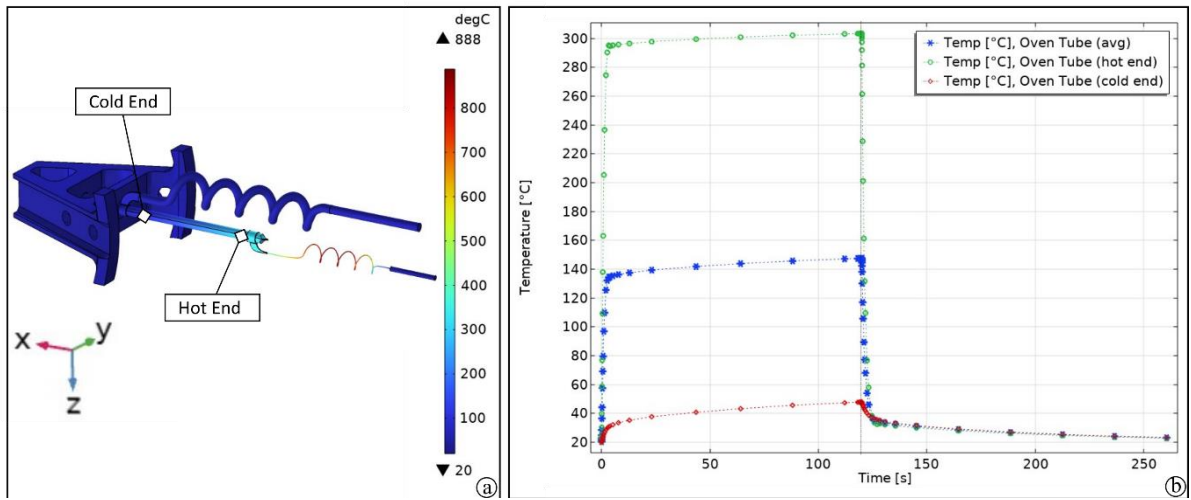

**Figure 18.** Mixed diameters electrical wires configuration: (a) temperature distribution in the mini oven at the peak heating and (b) temperature profile in relation to the application of electrical current

As a result of this study, the design shown in section (ii) was selected, as it allows faster response times and represents an improvement in term of heating efficiency (also regarding the temperature distribution) compared to the design discussed in sections (i) and (iii).

Following the analysis discussed above, prototyping activities on the oven were performed to develop an effective assembly procedure and test the oven functionality. With these, it was confirmed that no clogging occurs due to strontium vapour condensation, with no visible reduction in the tube inner diameter. This can be directly related to the small quantity of strontium used in the oven (<1000 µg), which is located mainly at the rear (hot) end of the tube.

## (c) Thermal cycling modelling

A thermal model was developed to simulate the conditions expected during future non-operational thermal vacuum tests, studying the temperature change for different regions of the assembly in relation to a cyclic temperature variation of the system's outer surface in contact with the thermal platen of a thermal vacuum test chamber. The conductive heat transfer between the platen and the assembly was simulated in an ideal scenario where a perfect thermal contact is established between the mating surfaces. The effect of radiative heat transfer of the trap elements towards the environment was also included in the model. This study aimed at understanding the temperature profile and heating/cooling rates of the system (probing the components close to the trapping region) to inform the thermal vacuum tests that will be conducted at more advanced stages in the project.

Several temperature cycles were simulated, each with specific values for the start/end temperature of the platen and a defined heating/cooling rate. These values and other test requirements were extracted in accordance with the ECSS standards [35, 38], with reference to the USA standards [39, 40]. Qualification test levels and durations were applied to the thermal vacuum test parameters, aiming at simulating the conditions encountered by a qualification model of the ion trap assembly. The values of the key parameters used in the developed thermal model are summarised in Table 7.

**Table 7.** Thermal model parameters summary

| Parameter | Value | Source |
|---|---|---|
| Temperature tolerance | ±2 °C | SMC-S-016 [40] |
| Temperature stability | ΔT <3 K/hr over a period of 30 minutes | SMC-S-016 [40] |
| Temperature range | [-50; 70] °C | ECSS-E-ST-10-03C [35] |
| Number of thermal cycles | 8 | ECSS-E-ST-10-03C [35] |
| Temperature transition rate | 20 °C /min | ECSS-E-ST-10-03C [35] |
| Minimum dwell time | 4 hours | SMC-S-016 [40] |

The predicted heating/cooling rates of the ion trap were extracted from the model, obtaining information about the time to reach steady-state conditions (temperature stability within tolerances, Table 7). An example of the plots obtained from the simulation is shown in Figure 19, for the specific case of heating the sub-assembly from 20 °C to 110 °C, for different regions of the sub-assembly.

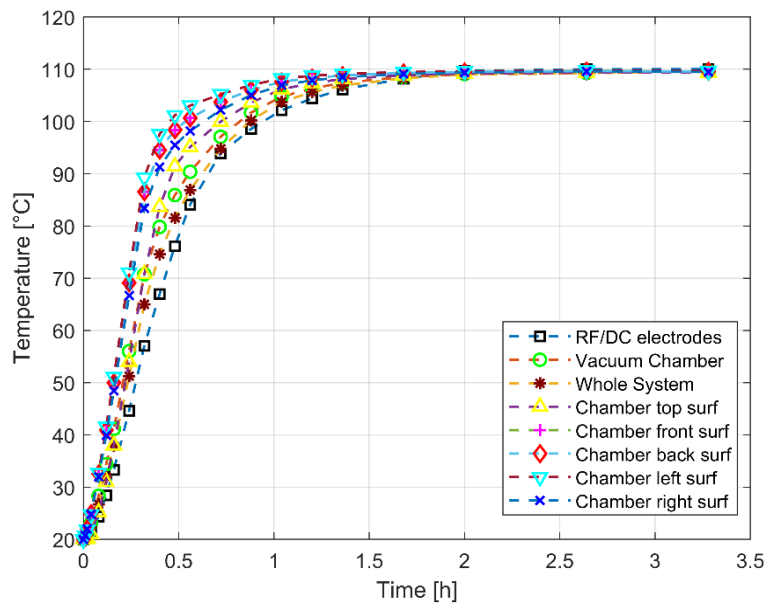

**Figure 19.** Temperature profile corresponding to different locations of the ion trap assembly, for a reference temperature increase from 20 °C to 110 °C. The dashed lines for each case represent fit curves to the predicted data points.

The data from the simulations was used to inform the thermal vacuum testing plan, as it allowed extracting information regarding the heating/cooling rate and the time to reach thermal stability for each cycle. Specifically, the latter was used to estimate the cycle duration, as that is combined with the minimum dwell time. With this value and the total number of cycles, the total duration of the test is predicted to be about 115 hours, which will be split across different testing sessions. In Figure 20, the results from the different simulations for every cycle are combined with the target temperature profile for the thermal vacuum test, showing the predicted total duration of the test and the temperature profile corresponding to the RF/DC electrodes (average value across the area of the probed components).

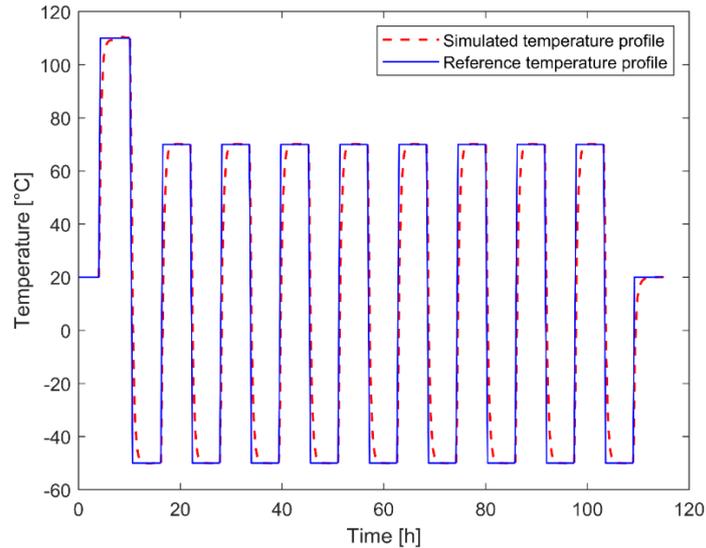

**Figure 20.** Thermal cycling: target temperature profile and results from simulation

During the non-operational thermal vacuum tests to be conducted in later stages of this project, the temperature of the trap will be monitored using in-vacuum PT100 sensors, that will also be used during the operation of the trap to apply the necessary correction for the total blackbody shift. These sensors will also contribute to an estimation of the environmental temperature in the region of the trap electrode structure during the operational phase, when the full application of various levels of active and passive thermal control are operational, thereby allowing the blackbody shift correction to be determined.

## 6. Conclusion

The design of the ion trap physics package system has been presented, discussing the results from the FEA of the ion trap in terms of both structural and thermal studies. The design addresses potential manufacturability and integration issues, introducing the modifications from the selected baseline [19].

The results of the structural studies (section 3) demonstrated that the design satisfies the mechanical strength requirements, showing that the ion trap assembly can withstand the loads associated with a space launch. The electrostatic studies (section 4) investigated the relationship between dimensional variations of the trap's key elements and their functionality. These concluded that geometrical tolerances play an important role in altering the trap pseudo-potential but did not suggest a revision of the nominal trap geometry that could lead to quantifiable improvements in the trapping efficiency. The thermal studies (section 5) allowed the investigation of the ion trap thermal behaviour from different perspectives. Specifically, the electro-thermal model of the whole system predicted the temperature distribution within the trap following the application of the applied electrical potential difference. The atomic mini-oven model enabled the exploration of different configurations of the proposed design and understanding of the relationship between the applied electric current and the temperature profile in the oven tube. A thermal cycling model of the ion trap was also presented, investigating the temperature profile (heat/cooling rates and time to steady-state conditions) as a function of a reference target temperature. This model simulated the conditions expected during thermal vacuum tests in preparation of the tests that will be conducted in future phases of the project.

Following the studies described in this manuscript, the manufacture and assembly of the integrated system will be undertaken, followed by in depth performance and thermal vacuum testing. Further development activities will be carried out in potential follow-ups to the project working towards an engineering qualification model.

Acknowledgments. We acknowledge the support of the European Space Agency for this work under contract 4000134936/21/NL/AR.